\documentclass{kluwer}
\usepackage{epsfig}

\newcommand{\be}{\begin{equation}}
\newcommand{\ee}{\end{equation}}
\newcommand{\beq}{\begin{eqnarray}}
\newcommand{\eeq}{\end{eqnarray}}

\begin{document}
\begin{article}
\begin{opening}

\title{GLOBAL CORONAL SEISMOLOGY}

\author{I. \surname{BALLAI}}

\runningauthor{BALLAI} \runningtitle{Global coronal seismology}

\institute{Solar Physics and Space Plasma Research Centre
(SP$^2$RC), Department of Applied Mathematics, University of
Sheffield, Hounsfield Road, Hicks Building, Sheffield, S3 7RH,
U.K. \email{i.ballai@sheffield.ac.uk}
             }

\date{Received ; accepted }

\begin{abstract}

Following the observation and analysis of large-scale coronal
wave-like disturbances, we discuss the theoretical progress made
in the field of {\it global coronal seismology}. Using simple
mathematical techniques we determine average values for magnetic
field together with a magnetic map of the quiet Sun. The
interaction between global coronal waves and coronal loops allows
us to study loop oscillations in a much wider context, {\it i.e.}
we connect global and local coronal oscillations.

\end{abstract}

\end{opening}

\section{Introduction}

The possibility of using waves propagating in solar atmospheric
plasmas to infer quantities impossible to measure (magnetic field,
transport coefficients, fine structuring, {\it etc.}) became a
reality after high cadence observation of oscillatory motion was
made possible by space and ground-based telescopes. These
observations combined with theoretical models allow to develop a
new branch of solar physics called {\it coronal seismology}.
Pioneering studies by Uchida (1970), Roberts, Edwin and Benz
(1984), Aschwanden {\it et al.} (1999) and Nakariakov {\it et al.}
(1999), have formed the basis of a very promising and exciting
field of solar physics.

Traditionally, the terminology of coronal seismology was used
mainly to describe the techniques involving waves propagating in
coronal loops. Since then, this word has acquired a much broader
significance and the technique is generalised to acquire
information about the magnetic solar atmosphere (De Pontieu,
Erd\'elyi and James, 2004; Erd\'elyi, 2006). Coronal seismology
uses waves which are localized to a particular magnetic
structures, therefore it would be necessary to label these seismic
studies as {\it local coronal seismology}. After the discovery of
large-scale wave-like disturbances, such as EIT waves, X-ray
waves, {\it etc.}, it became necessary to introduce a new
terminology, {\it i.e.} {\it global coronal seismology} where the
information is provided by global waves propagating over very
large distances, sometimes comparable to the solar radius.
Although this may seem a separate subject, in reality these two
aspects of coronal seismology are very much linked. A global wave
generated by sudden energy releases (flares, CMEs) can interact
with active region loops or prominences and localized loop or
prominence waves and oscillations are emerging so, there must be a
link between the generating source and flare-induced waves in
coronal loops.

Global waves have been known since the early $1960s$. Although it
is still not known how the release of energy and energized
particles will transform into waves, today it is widely accepted
that these disturbances are similar to the circularly expanding
bubble-like shocks after atomic bomb explosion or shock waves
which follows the explosion of a supernova. Thanks to the
available observational facilities, global waves were observed in
a range of wavelengths in different layers of the solar
atmosphere. A pressure pulse can generate seismic waves in the
solar photosphere propagating with speeds of 200\,--\,300
kms$^{-1}$ (Kosovichev and Zharkova, 1998; Donea {\it et al.},
2006). Higher up, a flare generates very fast super-alfv\'enic
shock waves known as {\it Moreton waves} (Moreton and Ramsey,
1960), best seen in the wings of $H{\alpha}$ images, propagating
with speeds of 1000\,--\,2000 km$s^{-1}$. In the corona, a flare
or CME can generate an EIT wave (Thompson {\it et al.}, 1999)
first seen by the SOHO/EIT instrument or an X-ray wave seen in SXT
(Narukage {\it et al.}, 2002). There is still a vigourous debate
how this variety of global waves are connected (if they are, at
all). Co-spatial and co-temporal investigations of various global
waves have been carried out but without a final widely accepted
result being reached. The present study deals with the properties
of EIT waves, therefore some characteristics of these waves will
be given below.

Unambiguous evidence for large-scale coronal impulses initiated
during the early stage of a flare and/or CME has been provided by
the Extreme-ultraviolet Imaging Telescope (EIT) observations
onboard SOHO and by TRACE/EUV. EIT waves propagate in the quiet
Sun with speeds of 250\,--\,400 km s$^{-1}$ at an almost constant
altitude. At a later stage in their propagation EIT waves can be
considered a freely propagating wavefront which is observed to
interact with coronal loops (see, {\it e.g.} Wills-Davey and
Thompson, 1999). Using TRACE/EUV 195 \AA\ observations, Ballai,
Erd\'elyi and Pint\'er (2005) have shown that EIT waves (seen in
this wavelength) are waves with average periods of the order of
400 seconds. Since at this height, the magnetic field can be
considered vertical, EIT waves were interpreted as fast MHD waves.

\section{Coronal Global EIT Waves and their Applications}

The observations of EIT waves propagating in the solar corona
allowed us to shed light on some elementary properties of coronal
global EIT waves, however, the available observational precision
does not permit us yet to determine more characteristics of these
waves.

One of the un-answered problems related to EIT waves is connected
to their propagation. The core of the problem resides in the lack
of detection of an EIT wave with every flare or CME. This could be
explained partly by the poor temporal resolution of the SOHO
satellite (the only satellite giving full-disk EUV images at the
moment) where frames are available with a low cadence, therefore
EIT waves generated near the limb simply cannot be recorded. In
general, EIT waves seen by SOHO are generated by sources which are
located near the centre of the solar disk. EUV images provided by
TRACE are much better to use, although the field of view of this
instrument is limited. Since EIT waves propagate over a large area
of the solar surface (at a certain altitude) they are dispersive.
Other ingredients to be considered are the stratification of the
medium and the inhomogeneous character of the plasma. All of these
factors influence the propagation of coronal EIT waves.

Another plausible explanation for the absence of EIT waves
associated with every flare or CME might be that EIT waves diffuse
very rapidly, {\it i.e.} they become evanescent in a short time
after their launch. This means that only those EIT waves could be
observed which propagate as guided (trapped) waves. The MHD
equations in a gravitationally-stratified plasma allows as a
solution the magnetoacoustic/magnetogravitational waves of growing
amplitude with time (upward propagating waves) and decreasing
amplitude with time (downward propagating waves). Trapped EIT
waves might arise as a combination of magnetoacoustic and
magnetogravitational waves propagating in opposite directions.
Further investigations of the possibility of trapping spherical
waves in a dissipative medium are needed.

\subsection{Interaction of EIT waves with other coronal magnetic entities}

In this subsection we enumerate a few possible phenomena arising
from the interaction of global EIT waves with coronal magnetic
entities. According to the classical picture, EIT waves collide
with coronal loops resulting in a multitude of modes generated in
loops either in the form of standing oscillations or propagating
waves. Both types of waves have the general property that they
decay very rapidly in a few wavelengths or periods (see {\it e.g.}
Nakariakov {\it et al.}, 1999; Aschwanden {\it et al.}, 2002).
This damping was later used to diagnose the magnetic field inside
coronal loops (Nakariakov {\it et al.}, 1999), transport
coefficients for slow waves or global fast waves, sub-structuring,
heating function, {\it etc.}

In coronal loops we consider only the transversal generation of
waves, {\it i.e.} waves and oscillations are triggered by the
interaction of EIT waves and coronal loops. From the EIT wave
point of view, a coronal loop (similar to an active region or
coronal hole) is an entity with a stronger magnetic field (at
least one order of magnitude) than the medium in which they
propagate (quite Sun). Therefore, beside transferring energy to
coronal loops, EIT waves can be scattered, reflected, and
refracted (Terradas and Ofman, 2004). Without claiming
completeness, we can draw a few
conditions that could influence the appearance of coronal loop oscillations:\\
- {\it height of the loop}: since EIT waves propagate at certain
heights in the solar corona, it is likely that not all loops will
interact with the global waves. Schrijver, Aschwanden and Title
(2002) pointed out that only those loops will be affected by EIT
waves whose
heights exceed 60\,--\,150 Mm. This means that cool, low-lying loops will not interact with EIT waves.\\
- {\it the height of the interaction between EIT waves and coronal
loops}: this factor simply means that it is easier to generate
oscillations in a loop if the interaction point between the EIT
wave and coronal loop is closer to the apex of the loop rather
than the footpoint.\\
- {\it Orientation of the loop}: if the front of the EIT wave is
perpendicular to the plane of the coronal loop the interaction
between the EIT waves and the coronal loop occurs in two points at
the same time. If the loop is stiff enough, a standing oscillation
can be easily excited. If the front is not perpendicular, the
collision between the EIT wave and loops occurs in two points
delayed in time by $\tau=s\cos\alpha/v_{EIT}$, where $s$ is the
distance between the footpoints, $\alpha$ is the attack angle, and
$v_{EIT}$ is the propagation speed of the EIT wave. In this case,
standing modes can be excited only in very special cases. Another
important element is the orientation
of the coronal loop with respect to the vertical axis (inclination).\\
- {\it distance between the flaring site and coronal loop} (or
energy of EIT waves): During their propagation, EIT waves are
losing energy due to the geometrical damping (dilatation of the
front) and due to some physical damping effects. Therefore it
might happen that the energy of an EIT wave originating from a
distant flare is not enough to
dislocate the loop.\\
- {\it radius of the loop and the density contrast (or Alfv\'en
speed contrast)}: a massive loop is much harder to dislocate than
a thin loop. The ratio between the densities in the loop and its
environment is known to influence the amplitude of oscillations.

In order to describe quantitatively the interaction between EIT
waves and coronal loops, we suppose a medium in which the coronal
loop is situated, for simplicity, in a magnetic-free medium (in
fact this constraint can be relaxed and the result is obvious)
retains its identity and does not disperse or fragment. The tube
is considered thin, {\it i.e.} its radius is small relative to
other geometrical scales of the problem. During the wave
propagation we suppose a quasi-static pressure balance to be
maintained at all times.

An EIT wave colliding with a coronal loop exerts a force which
will need to work against two forces, one being the elastic force
of the tube represented by the magnetic tension of the tube and
inertia of the fluid element which needs to be displaced.

The equilibrium of the tube is prescribed by the hydrostatic
equilibrium where pressure forces are in equilibrium with the
gravitational force and the lateral pressure balance
$p_i+B_i^2/2\mu=p_e$ is satisfied, with $p_i$ and $p_e$ being the
kinetic (thermal) pressure inside the tube and the environment,
$B_i$ the interior magnetic field  and $g$ is the gravitational
acceleration at the solar surface. If we denote by $\rho_i$ and
$\rho_e$ the locally homogeneous densities inside and outside the
tube and $v_A(=B_i/(\mu\rho_i)^{1/2})$ the Alfv\'en speed, then
the equation describing the variation of displacement of the fluid
element, $\xi(z,t)$, is (a similar equation has been obtained by
Ryutov and Ryutova (1975) in a different context)
\begin{equation}
\frac{\partial^2\xi}{\partial
t^2}=-g\frac{\rho_i-\rho_e}{\rho_i+\rho_e}\frac{\partial
\xi}{\partial
z}+\frac{\rho_i}{\rho_i+\rho_e}v_A^2\frac{\partial^2\xi}{\partial
z^2}. \label{eq:2.1}
\end{equation}
Let us introduce a new variable such that
$\xi(z,t)=Q(z,t)\exp(\lambda z)$, where the value of $\lambda$ is
chosen such that the first-order derivatives with respect to the
coordinate $z$ vanish. After a straightforward calculation we
obtain that the dynamics of generated waves in the coronal loop as
a result of the interaction of a global wave with coronal loop is
described by
\[
\frac{\partial^2Q}{\partial t^2}-c_K^2\frac{\partial^2 Q}{\partial
z^2}+\omega_C^2Q=0, \quad
\omega_C=\frac{g(d-1)}{2v_A\sqrt{d(d+1)}},
\]
\begin{equation}
\quad c_K=v_A\left(\frac{d}{1+d}\right)^{1/2}\label{eq:2.2}
\end{equation}
with $d=\rho_i/\rho_e$ being the filling factor. Equation
(\ref{eq:2.2}) is the well-known Klein-Gordon (KG) equation
derived and studied earlier in solar MHD wave context by, {\it
e.g.} Rae and Roberts (1982), Hargreaves (2005), Ballai, Erd\'elyi
and Hargreaves (2006). The quantity $c_K$ is the kink speed of
waves and it is regarded as a density-weighted Alfv\'en speed. The
coefficient $\omega_C$ is the cut-off frequency of waves and is a
constant quantity for an isothermal medium.

The waves corresponding to the Eq. (\ref{eq:2.2}) are dispersive,
{\it i.e.} waves with smaller wavelength (larger $k$) propagating
faster. Waves with smaller wave number will have smaller group
speed, the maximum of the group speed (at $k\rightarrow \infty$)
being $c_K$. Another essential property of the KG equation is that
it describes waves which are able to propagate if their frequency
is larger than the cut-off frequency. For typical coronal
conditions ($v_A$=1000 km s$^{-1}$, $d=10$) we obtain that waves
will propagate if their frequency is greater than 0.11 mHz or
their period is smaller than 150 minutes.

For simplicity, let us suppose that the fast kink mode in the
coronal loop is generated by the interaction of an EIT wave with a
loop and the forcing term of the interaction is modelled by a
delta-pulse, {\it i.e.} the equation describing the dynamics of
impulsively generated fast kink mode is given by
\begin{equation}
\frac{\partial^2Q}{\partial t^2}-c_K^2\frac{\partial^2 Q}{\partial
z^2}+\omega_C^2Q=\delta(z)\delta(t). \label{eq:2.6}
\end{equation}
This equation can be solved using standard Laplace transform
technique to yield
\begin{equation}
Q(z,t)=\frac{c_K}{2}J_0\left(\omega_C\sqrt{c_K^2t^2-z^2}\right)H(c_Kt-|z|),
\label{eq:7}
\end{equation}
where $J_0(z)$ is the zeroth-order Bessel function and $H(z)$ is
the Heaviside function. The impulsive excitation of waves in a
flux tube leads to the formation of a pulse that propagates away
with the speed $c_K$, followed by a wake in which the flux tube
oscillates with the frequency $\omega_C$. A typical temporal
variation of the amplitude of kink waves (keeping the height
constant) would show that the amplitude of the mode decreases
(even in the absence of dissipation) and an e-fold decay occurs in
about 400 seconds .

Recently Terradas, Oliver and Ballester (2005) have studied the
interaction between the coronal loops and EIT waves in the
zero-beta limit considering a spatial initial condition. They
obtained that the generated oscillations in the coronal loop decay
asymptotically as $t^{-1/2}$. Kink oscillations are weakly
affected by dissipation, therefore the consideration of any
non-ideal effect to supplement the KG equation would not lead to a
significant change. It is accepted that damping due to the
resonant absorption could explain the damping of kink oscillations
in coronal loops (Ruderman and Roberts, 2002; Goossens, Andries
and Aschwanden, 2002).

It can be shown that the consideration of resonant absorption as a
damping mechanism in the governing equation leads to a similar
equation we would obtain taking into account dissipation. Ballai,
Erd\'elyi and Hargreaves (2006) showed that the evolution equation
is modified by an extra term forming a Klein-Gordon-Burgers (KGB)
equation
\begin{equation}
\frac{\partial^2Q}{\partial t^2}-c_K^2\frac{\partial^2 Q}{\partial
z^2}+\omega_C^2Q-\nu\frac{\partial^3Q}{\partial z^2\partial t}=0,
\label{eq:2.8}
\end{equation}
where $\nu$ is a coefficient which could play the role of any
dissipative mechanism or a factor which include the damping due to
resonant absorption (in fact $\nu$ is inversely proportional to
the gradient of Alfv\'en speed) and would describe the transfer of
energy from large to small scales (see, {\it e.g.} Ruderman and
Goossens, 1993).

Waves in this approximation can have a temporal (keeping $k$ real)
and spatial damping (keeping $\omega$ real), the decay rate and
length, supposing the ansatz $Q(z,t)\sim \exp[i(\omega t-kz)]$,
are given by
\begin{equation}
\omega_i=\frac{i\nu k^2}{2}, \quad k_i\approx
-\frac{\nu\omega}{2c_K}\sqrt{\frac{\omega^2-\omega_C^2}{c_K^4+\nu^2\omega^2}}.\label{eq:2.9}
\end{equation}
The KGB equation can be solved using initial/boundary conditions
to describe the evolution of kink modes for different kind of
sources, {\it e.g.} monochromatic source ($A(t)=V_0e^{i\Omega
t}$), delta-function pulse ($A(t)=V_0\delta(\omega_Ct/2\pi)$),
{\it etc.} using numerical methods. Asymptotic analysis ($t\gg
z/c_K)$ shows that these waves decay as $t^{-3/2}$ (Ballai,
Erd\'elyi and Hargreaves, 2006).

Another important factor is the energy of EIT waves. Recently
Ballai, Erd\'elyi and Pint\'er (2005), using a simple energy
conservation, found the minimum energy an EIT wave should have to
produce a loop oscillation. Using their results we studied a few
loop oscillation events presented by Aschwanden {\it et al.}
(2002) and the minimum energy of EIT waves necessary to produce
the observed oscillations are shown in Table 1. The geometrical
size of loops and the number densities given by Aschwanden {\it et
al.} (2002) have been used.\\
\begin{table}

\begin{tabular}{clllc}
  \hline
  Date(yyyymmdd) & L(Mm) & R(Mm) & n($\times 10^8$ cm$^{-3}$) & E(J) \\
  \hline
  1998 Jul 14 & 168 & 7.2 & 5.7 & $2.2\times 10^{17}$ \\
  1998 Jul 14 & 204 & 7.9 & 6.2 & $9.7\times 10^{18}$ \\
  1998 Nov 23 & 190 & 16.8 & 3 & $1.3\times 10^{19}$ \\
  1999 Jul 04 & 258 & 7 & 6.3 & $3.9\times 10^{16}$ \\
  1999 Oct 25 & 166 & 6.3 & 7.2 & $1.6\times 10^{18}$ \\
  2000 Mar 23 & 198 & 8.8 & 17 & $5.2\times 10^{16}$ \\
  2000 Apr 12 & 78 & 6.8 & 6.9 & $2.5\times 10^{16}$ \\
  2001 Mar 21 & 406 & 9.2 & 6.2 & $7.4\times 10^{16}$ \\
  2001 Mar 22 & 260 & 6.2 & 3.2 & $1.9\times 10^{16}$ \\
  2001 Apr 12 & 226 & 7 & 4.4 & $1.4\times 10^{18}$ \\
  2001 Apr 15 & 256 & 8.5 & 5.1 & $1.4\times 10^{16}$ \\
  2001 May 13 & 182 & 11.4 & 4 & $2.2\times 10^{18}$ \\
  2001 May 15 & 192 & 6.9 & 2.7 & $1.6\times 10^{19}$ \\
  2001 Jun 15 & 146 & 15.8 & 3.2 & $1.1\times 10^{17}$ \\
  \hline
\end{tabular}
\caption[]{The minimum energy of EIT waves which could produce the
loop oscillations studied by Aschwanden {\it et al.} (2002).}
\end{table}
The obtained energies are in the range of $10^{16}-10^{19}$ J with
no particular correlation with the length and radius of the loop.
Similar to this approach we can estimate the minimum energy of an
EIT wave to produce a displacement of 1 pixel in TRACE/EUV 195
\AA\ images using the relation
$E=1.66\times10^6L\left(\rho_iR^2+\rho_e/\lambda_e^2\right),$ (J)
where $L$ and $R$ are the length and radius of the loop, and
$\lambda_e^{-1}$ the decay length of perturbations outside the
cylinder given by
\begin{equation}
\lambda_e^{2}=\frac{(c_{Se}^{2}-c_K^2) (v_{Ae}^{2}-c_K^2)}
{(c_{Se}^{2}+v_{Ae}^2)(c_{Te}^{2}-c_K^2)}k^2, \label{eq:14}
\end{equation}
with $c_{Te}$, $c_{Se}$, and $v_{Ae}$ being the cusp, sound and,
Alfv\'en speeds in the region outside the loop and $k$ is the
wavenumber. The energy range is in the interval $3\times
10^{17}-3\times 10^{18}$ J for loop lengths and radii varying in
the intervals 60\,--\,500 Mm and 1\,--\,10 Mm.

\subsection{Determination of magnetic field values}

Observations show that EIT waves propagate in every direction
almost isotropically on the solar disk, therefore we can
reasonably suppose that they are fast magnetoacoustic waves (FMWs)
propagating in the quiet Sun perpendicular to the vertical
equilibrium magnetic field. The representative intermediate line
formation temperature corresponding to the $195$ \AA\ wavelength
is $1.4\times 10^6$ K. The sound speed corresponding to this
temperature is $179$ km s$^{-1}$. Since the FMWs propagate
perpendicular to the field, their phase speed is approximated by
$(c_S^2+v_A^2)^{1/2}$.

The propagation height is an important parameter as a series of
physical quantities (density, temperature, {\it etc.}) in the
solar atmosphere have a height dependence. Given the present
status of research on the propagation of EIT waves, there is no
accepted value for the propagation height of these waves. For a
range of the plasma parameters we can derive average values for
the magnetic field by considering the propagational characters of
EIT waves. Therefore, we study the variation of various physical
quantities with respect to the propagation height of EIT waves.

We recall a simple atmospheric model developed by Sturrock,
Wheatland and Acton (1996). The temperature profile above a region
of the quiet Sun, where the magnetic field is radial, is given by
\begin{equation}
T(x)=\left[T_0^{7/2}+\frac{7R_{\odot}F_0}{2a}\left(1-\frac{1}{x}\right)\right]^{2/7}.
\label{eq:15}
\end{equation}
Here $F_0$ is the inward heat flux ($1.8\times 10^5$ erg cm$^{-2}$
s$^{-1}$), $x$ is the normalized height coordinate defined by
$x=r/R_{\odot}$, $T_0$ is the temperature at the base of the model
(considered to be $1.3\times 10^6$ K) and $a$ is the coefficient
of thermal conductivity. The quantity $a$ is weakly dependent on
pressure and atmospheric composition; for the solar corona a value
of $10^{-6}$ (in cgs units) is appropriate (Nowak and
Ulmschneider, 1977). Assuming a model atmosphere in hydrostatic
equilibrium we obtain that the number density, based on the
temperature profile supposed in Eq. (\ref{eq:15}), is
\begin{equation}
n(x)=\frac{n_0T_0}{T(x)}\exp[-\delta(T(x)^{5/2}-T_0^{5/2})], \quad
\delta=\frac{2\mu GM_{\odot}m_pa}{5k_BR_{\odot}^2F_0},
\label{eq:16}
\end{equation}
with $G$ the gravitational constant, $M_{\odot}$ the solar mass,
$k_B$ is the Boltzmann's constant; $\mu=0.6$ is the mean molecular
weight; $m_p$, proton mass and $n_0=3.6\times 10^8$ cm$^{-3}$ the
density at the base of corona. Having the variation of density
with height and the value of Alfv\'en speed deduced from the phase
speed of EIT waves, we can calculate the magnetic field using
$B=v_A(4\pi m_pn)^{1/2}$. Evaluating the relations above, the
variation of temperature, density, Alfv\'en speed and magnetic
field with height is shown in Table 2.
\begin{table}
\caption{The variation of the temperature (in MK), density (in
units of $10^8$ cm$^{-3}$), Alfv\'en and sound speeds (in units of
$10^7$ cm s$^{-1}$) and magnetic field (in G) with
 height above the photosphere
 for an EIT wave propagating with a speed of (a) 250 km s$^{-1}$, and
(b) 400 km s$^{-1}$, respectively.}
\begin{tabular}{cccccccc}
  \hline
  $r/R_{\odot}$ & T & n & $c_S$  & $v_A^{(a)}$
  & B$^{(a)}$ & $v_A^{(b)}$ & B$^{(b)}$ \\
  \hline
  1.00 & 1.30 & 3.60 & 1.72 & 1.81 & 1.57 & 3.61 & 3.13 \\
  1.02 & 1.41 & 3.30 & 1.80 & 1.73 & 1.44 & 3.57 & 2.97\\
  1.04 & 1.50 & 3.10 & 1.85 & 1.67 & 1.34 & 3.54 & 2.85 \\
  1.06 & 1.58 & 2.95 & 1.90 & 1.61 & 1.27 & 3.51 & 2.76\\
  1.08 & 1.64 & 2.83 & 1.94 & 1.57 & 1.21 & 3.49 & 2.69 \\
  1.10 & 1.70 & 2.73 & 1.97 & 1.52 & 1.15 & 3.47 & 2.63\\
  \hline
\end{tabular}
\end{table}
Two cases are derived for EIT waves propagating strictly
perpendicular to the radial magnetic field with a speed of (a)
$250$ km s$^{-1}$ and, (b) $400$ km s$^{-1}$, respectively. The
values of the physical quantities show some change for a given
propagation speed but will have little effect on the final
results.

For an average value of EIT wave speed of $300$ km s$^{-1}$
propagating at 0.05 $R_{\odot}$ above the photosphere we find that
the magnetic field is $1.8$ G. If we apply $Br^2=const.$, {\it
i.e.} the magnetic flux is constant, we find that at the
photospheric level the average magnetic field is $2.1$ G which
agrees well with the observed solar mean magnetic field (Chaplin
{\it et al.} 2003).
\begin{figure}
\begin{center}
  \psfig{file=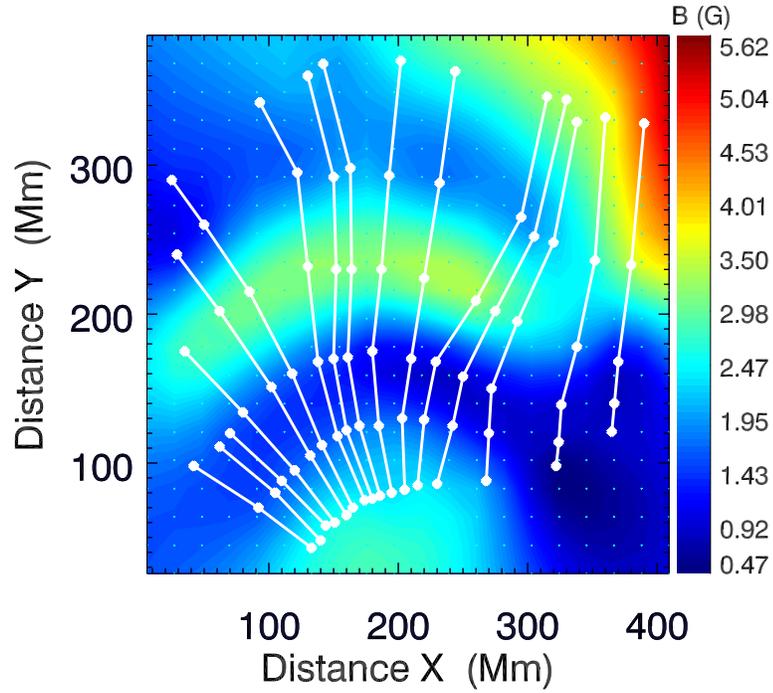,width=11.cm,clip=}
  \caption{Magnetic map of the quite Sun obtained using an EIT wave observed by TRACE/EUV in
  195 \AA\ . }
  \end{center}
\end{figure}
EIT waves considered as fast MHD waves can also be used to
determine the value of the radial component of the magnetic field
at every location allowed by the observational precision. In this
way, using the previously cited TRACE observations we can
construct a magnetic map of the quiet Sun (see Figure 1), in other
words EIT waves can serve as probes in a {\it magnetic tomography}
of the quiet Sun. If points are joined across the lines, we will
obtain the location of the EIT wavefront. Magnetic field varies
between 0.47 and 5.62 G, however, these particular values should
be handled with care as the interpolation will introduce spurious
values at the two ends of the interval. It should be noted that
this result has been obtained supposing a single value for
density, in reality both magnetic field and density can vary along
the propagation direction, as well. The method we employed to find
this magnetic map (magnetic field derived via the Alfv\'en speed)
means that density and magnetic field cannot be determined at the
same time. Further EUV density sensitive diagnostics line ratio
measurements are required to establish a density map of the quiet
Sun which will provide an accurate determination of the local
magnetic field.

In conclusion, EIT waves propagating in the solar corona exhibit a
wide range of applicabilities for plasma and field diagnostics.
The fact that during their propagation EIT waves cover a large
area of the solar surface (in the coronal) allows us to sample the
magnetic field in the quiet Sun. EIT waves could serve as a link
between eruptive events and localised oscillations, {\it e.g.}
loop oscillations could be studied in a much broader context.
Using a simple model we found that the minimum energy an EIT wave
should have to produce a detectable loop oscillation is in the
range of $10^{16}-10^{19}$ J.

Problems to be tackled in the future should include the study of
attenuation of EIT waves with the aim of providing information
about the magnitude of transport coefficients in the quiet Sun.
The old problem of connecting different global waves still remain
to be addressed.

Despite of the lack of high precision observations, EIT waves show
a great potential for magneto-seismology of the solar corona.

\acknowledgements The author acknowledges the financial support
offered by the Nuffield Foundation (NUF-NAL 04) and NFS Hungary
(OTKA, T043741). The help by B. Pint\'er and M. Douglas is
appreciated.

\end{article}
\end{document}